\documentclass[11pt]{article}

\def\problembox#1{\vspace{2mm}\noindent\fbox{\begin{minipage}{0.98\linewidth}#1
 \end{minipage}}\vspace{2mm}\noindent }

\usepackage[a4paper,margin=27mm]{geometry}
\usepackage[T1]{fontenc}
\usepackage[utf8]{inputenc}
\usepackage{lmodern}
\usepackage{microtype}
\usepackage{amsmath,amssymb,amsthm}
\usepackage{mathtools}
\usepackage{algorithm}
\usepackage{algpseudocode}
\usepackage{booktabs}
\usepackage{longtable}
\usepackage{xcolor}
\usepackage{float}
\usepackage{tikz}
\usetikzlibrary{arrows.meta,calc}
\usepackage[hidelinks]{hyperref}

\definecolor{figStatic}{HTML}{DCEAF7}
\definecolor{figInsert}{HTML}{FBE0AB}
\definecolor{figExcluded}{HTML}{DFE1E4}
\definecolor{figMarker}{HTML}{B63843}
\definecolor{figOutput}{HTML}{DCEEDB}
\tikzset{
  figcell/.style={draw=black!60,minimum width=6.8mm,minimum height=6.8mm,
                  inner sep=0pt,font=\small},
  figlabel/.style={font=\footnotesize,align=center},
  figarrow/.style={-{Stealth[length=2mm]},thick},
}

\newcommand{\BWT}{\operatorname{BWT}}
\newcommand{\BBWT}{\operatorname{BBWT}}
\newcommand{\fnRank}{\mathsf{rank}}
\newcommand{\fnSelect}{\mathsf{select}}
\newcommand{\fnCount}{\mathsf{count}}
\newcommand{\LF}{\operatorname{LF}}
\newcommand{\FL}{\operatorname{FL}}
\newcommand{\myFunction}[1]{\textsf{#1}}

\newcommand*{\myblock}[1]{\noindent{\textbf{#1.~}}}

\title{Revisiting (Bijective) Burrows--Wheeler Transform Space--Time Trade-Off Constructions}
\author{Dominik K\"oppl and Tomohiro I}
\date{}

\usepackage[natbib,cref,theorems]{nikis}
\begin{document}
\maketitle

\begin{abstract}
The bijective Burrows--Wheeler transform (BBWT) replaces the delimiter of the ordinary BWT by a Lyndon factorization and sorts factor conjugates in \(\omega\)-order.  
It admits quadratic-time constant-workspace algorithms and linear-time linear-space algorithms.

We extend the small-space BWT trade-off of Crochemore et al.~[JDA,'15] to the BBWT and accelerate both constructions.
For \(\tau \in [3..n]\), 
let \(\sigma_\tau\le\min(\sigma,\tau+1)\) be the maximum construction-batch alphabet size.  
We construct either transform in
\( O\!\left( \frac{n^2}{\tau} \frac{\log\tau}{\log\log\tau} \right) \)
time and 
\(O(\tau\sigma_\tau)\) additional space, 
assuming \(\tau\sigma_\tau=O(n)\).  
This improves the transferred \(O((n^2/\tau)\log\tau)\) bound by a \(\log\log\tau\) factor.
If \(\sigma_\tau\in w^{O(1)}\), where \(w\in\Omega(\log n)\)
is the word size, the time improves to
\( O\!\left( \frac{n^2}{\tau} + \frac{n\log\tau}{\log\log\tau} \right). \)
For inversion, the same bounds hold with the global alphabet size
\(\sigma\) in place of \(\sigma_\tau\).
\end{abstract}

\section{Introduction}
\label{sec:introduction}

The Burrows--Wheeler transform (BWT)~\cite{burrows94bwt} is central to text compression and
indexing.  Its bijective variant (BBWT), introduced by Gil and Scott
(see \cite{bannai25survey}), needs no delimiter: it sorts the
conjugates of the text's Lyndon factors in \(\omega\)-order.
Construction must therefore respect both factor boundaries and cyclic order.
Linear-time, linear-space constructions are known for both
transforms~\cite{nong09sais,bannai24constructing,olbrich24generic}.

The in-place algorithm of \citet{koppl20inplace} processes Lyndon factors
from left to right, inserting each factor's letters from right to left.
We realize this order by reversing each factor in place.
Except for the first letter of a factor, each new letter is inserted
immediately after the position returned by LF.
Naive LF queries give \(O(n^2)\) time and \(O(1)\) workspace.

For the ordinary BWT, \citet[Thm.~3]{crochemore15bwt} give a trade-off:
for \(k\le n\), construction and inversion take
\(O\!\left((n^2/k+n)\log k\right)\) time in the comparison model using \(O(k\tilde{\sigma}_k)\) additional memory cells, where
\(\tilde{\sigma}_k\le\min(|\Sigma|,k)\) denotes the maximum number of distinct letters in any length-\(k\) substring of the input text.
\citet{louza17bwt} extended this trade-off to simultaneous construction
of the BWT and LCP array without changing the asymptotic bounds.
We extend the construction trade-off to the BBWT
(\cref{thm:bbwt:basicTradeoff}) and improve both constructions
(\cref{thm:bwt:btree}).  Our construction parameter
\(\sigma_\tau\) counts letters in a prepared batch, which need not be
contiguous in the original text; \cref{lem:batchintervals} relates it
to \(\tilde{\sigma}_\tau\).
For inversion we use the global alphabet size \(\sigma\).

Other BWT construction algorithms use different space models:
\citet{kempa19synchronizing} obtain sublinear time on packed binary strings,
while \citet{olbrich25fast} construct the BWT from Lyndon grammars of
repetitive texts.  Neither uses our in-place letter-cell model.

The key observation is that the Lyndon factorization changes only the insertion schedule.  
Once the schedule is fixed, the BBWT construction uses the same insertions 
and LF queries as the BWT construction, 
so both admit the same batch-and-breakpoint structure (BBS) representation~\cite{crochemore15bwt}.

The BBS resembles a bounded \emph{piece table}~\cite{newman12design,crowley98textsequences} of general text editors:
the piece table keeps the current transform static and records pending one-letter insertions separately.  
Unlike a general text editor, it stores at most \(\tau\) pending edits, 
materializes them after each batch, and maintains the letter counts needed for rank and LF\@.

\paragraph{Useful parameter range.}
To get sublinear space we need the condition that \(\tau\sigma_\tau=o(n)\).
Since \(\sigma_\tau\le\min(\sigma,\tau+1)\), 
the stronger condition \(\tau\min(\sigma,\tau)=o(n)\) suffices in the worst case; 
it is not necessary for texts with smaller batch alphabets.
For \(\tau=\omega(\sqrt n)\), sublinear space in particular requires \(\sigma_\tau=o(n/\tau)\).

For smaller \(\tau\), even a sorted array of pending breakpoints suffices:
if each update takes \(O(\tau)\) and scanning a static letter costs \(t_{\mathrm{lookup}}\),
the total time is \(O((n^2/\tau)t_{\mathrm{lookup}}+n\tau)\).
Thus, \(\tau=O(\sqrt n)\) makes the \(n\tau\) term negligible.

\section{Preliminaries}
\label{sec:preliminaries}

Let \(T = T[1..n]\) be a text over an ordered alphabet \(\Sigma\) of size $\sigma$.
Let $\$$ be a special letter that is smaller than any letter in \(\Sigma\) and does not occur in \(T\).
For the ordinary BWT, we use the end-marked transform \(\BWT(T\$)\).
As it has \(n+1\) letters, its unique delimiter is represented implicitly by its position, stored in \(O(1)\) additional machine words, while the \(n\) ordinary letters occupy the working array.
Inversion of the ordinary BWT takes this position as part of its input.
For a string \(S\), let \(|S|\) denote its length and let \(\Gamma(S)\) denote the set of distinct letters occurring in \(S\).
We write individual letters, such as \(\mathtt{c}\) and
\(\mathtt{d}\), in typewriter font to distinguish them from
numerical variables and indices.

A nonempty string \(S\) is a Lyndon word if it is strictly lexicographically smaller than all its nontrivial conjugates.
Every text \(T\) has a unique Lyndon factorization
\(
	T = T_1T_2\cdots T_\ell
\)
such that \(T_1 \ge T_2 \ge \cdots \ge T_\ell\) in lexicographic order and each \(T_i\) is a Lyndon word.
Duval's algorithm computes this factorization in linear time while outputting the factors from left to right.

The BBWT of \(T\) is obtained by considering all conjugates of all Lyndon factors \(T_1,\ldots,T_\ell\), sorting them in \emph{\(\omega\)-order}, and writing the letter that precedes each conjugate inside its factor.
For nonempty strings \(u,v\), this order compares the infinite periodic words \(u^\omega\) and \(v^\omega\) lexicographically; equal infinite words may be ordered consistently in either way when the resulting output letters coincide.
Equivalently, the BBWT is the extended Burrows--Wheeler
transform~\cite{mantaci07ebwt} of the multiset of Lyndon factors.

For a transform string \(L\), let
\( C_L[\mathtt{c}] := |\{i : L[i] < \mathtt{c}\}| \)
be the number of letters in \(L\) that are smaller than \(\mathtt{c}\).
Throughout the paper, all string positions are one-based.
We use the inclusive rank convention
\( L.\fnRank_{\mathtt{c}}(i) := |\{h\in[1..i] : L[h]=\mathtt{c}\}|, \)
with \(\fnRank_{\mathtt{c}}(L,0)=0\).
We also write \(\fnSelect_{\mathtt{c}}(L,q)\) for the position of the \(q\)-th occurrence of \(\mathtt{c}\) in \(L\).
The LF mapping is
\( \LF(i) := C[L[i]] + L.\fnRank_{L[i]}(i). \)
This formula only requires cumulative letter counts and rank queries on the current transform string.

For inversion, we use the forward mapping
\( \FL(i) := L.\fnSelect_{F[i]}(F.\fnRank_{F[i]}(i)), \)
where \(F\) is the conceptual string obtained by sorting the letters in \(L\).
We do not store the full array \(F\).
Instead, we access letter \(F[i]\) by $C_L$ as the unique letter \(\mathtt{c}\) whose cumulative interval contains \(i\),
i.e., $F[i] = c$ if and only if \(C_L[\mathtt{c}] < i \le C_L[\mathtt{c}+1]\).

Our algorithms work with a fixed working array \(X[1..n]\) of \(n\) \emph{letter cells}, 
each capable of holding one input letter or its rank-compressed code.
When the alphabet is already encoded as \([1..\sigma]\), these cells
can use \(O(n\log(\sigma+1))\) bits in total; 
if the original alphabet consists of arbitrary word-sized integer keys, the given array may instead occupy \(n\) words.
For construction, \(X\) initially contains the input text, i.e., \(X=T[1..n]\).  
For inversion, \(X\) initially contains the \(n\) ordinary letters of the transform to be inverted;
the delimiter position is additionally supplied for the ordinary BWT, while the BBWT needs no delimiter.
The \(n\) input/output cells of \(X\) are not counted as additional
space; all space bounds below count only auxiliary storage besides
this array.

We work on a RAM with constant-time random access to $X$
and constant-time arithmetic on indices and counters.
Letters can be compared in constant time.
For the basic bounds, ordered dictionaries are implemented
by comparison-based balanced search trees.
The acceleration in \cref{sec:btree} additionally uses
word-RAM operations.

\section{Main Result}
\label{sec:main-result}

The case \(\tau=1\) is exactly the constant-workspace in-place algorithm~\cite{koppl20inplace}.
We therefore assume throughout this section that \(\tau\) is an integer trade-off parameter 
with \(\tau \in [2..n]\).

The BBS of \citet{crochemore15bwt} stores the current transform
statically in a prefix of \(X\) and batches at most \(\tau\) updates.
Each pending insertion is a \emph{breakpoint}; the structure answers
queries on the logical, updated transform without shifting the static
part until the batch is flushed.

A \emph{construction batch}~\(B\) is a consecutive block of \(r\in[1..\tau]\) insertion steps in the prescribed (B)BWT construction order. 
Its source letters are the \(r\) letters consumed by these steps, 
prepared in insertion order in \(X[z+1..z+r]\), where \(z\) is the length of the static transform at the start of the batch.

The static part of length \(z\) is sampled into \(\tau\) segments of
length at most \(\lceil z/\tau\rceil\), allowing a segment scan in \(O(z/\tau+1)\) time.
The alphabet \(\Gamma_B\) of a construction batch \(B\) consists of
its at most \(\tau\) source letters and, when the batch begins inside a Lyndon factor, 
the letter needed for its first LF query.
Put \(\sigma_\tau:=\max_B|\Gamma_B|\); then \(\sigma_\tau\le\min(\sigma,\tau+1)\).
Unlike the substrings defining \(\tilde{\sigma}_k\), 
a construction batch draws letters from reversed Lyndon factors and need not be contiguous in \(T\).
However, we can still bound $\sigma_\tau$ in terms of \(\tilde{\sigma}_\tau\):

\begin{lemma}\label{lem:batchintervals}
The source letters consumed by one construction batch occupy the union of at most three intervals of \(T\), of total length at most \(\tau\).
Consequently, \(\sigma_\tau\le 3\tilde{\sigma}_\tau+1\).
The stronger inequality \(\sigma_\tau\le\tilde{\sigma}_\tau\) need not hold.
\end{lemma}

\begin{proof}
If the batch lies within one reversed factor, its source cells form one interval of \(T\).  
Otherwise, it consists of a prefix of its first
factor, zero or more complete intermediate factors, and a suffix of
its last factor: at most three intervals of total length \(r\le\tau\).
For their lengths \(r_1+r_2+r_3\le\tau\), monotonicity of
\(\tilde{\sigma}_\ell\) gives
\(|\Gamma(X[z+1..z+r])|
\le\sum_i\tilde{\sigma}_{r_i}
\le3\tilde{\sigma}_\tau\).
The boundary LF letter adds at most one.
\end{proof}

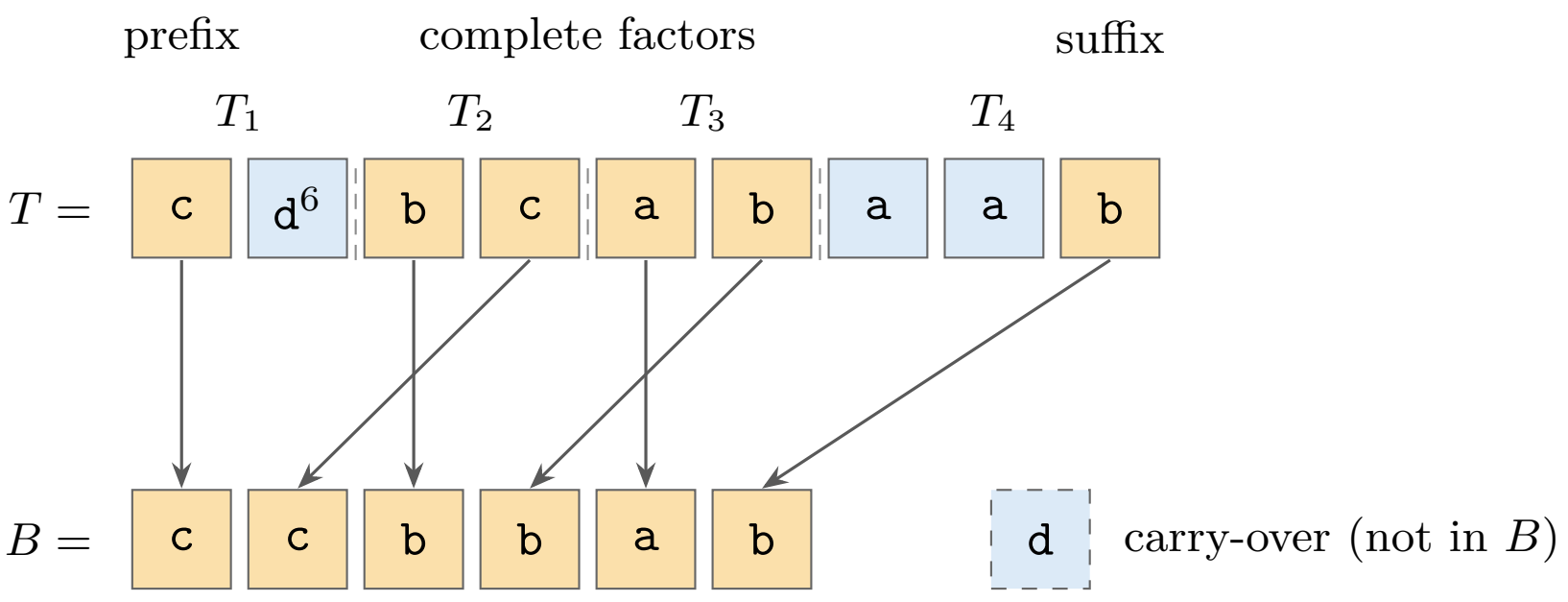
\begin{figure}[t]
\centering
\begin{tikzpicture}[x=0.80cm,y=0.76cm]
  \node[figlabel,anchor=east] at (-0.6,3) {\(T=\)};
  \node[figcell,fill=figInsert] at (0,3) {\(\mathtt c\)};
  \node[figcell,fill=figStatic] at (1,3) {\(\mathtt d^6\)};
  \foreach \i/\letter/\shade in {2/b/figInsert,3/c/figInsert,4/a/figInsert,5/b/figInsert,6/a/figStatic,7/a/figStatic,8/b/figInsert} {
    \node[figcell,fill=\shade] at (\i,3) {\(\mathtt{\letter}\)};
  }
  \foreach \x/\name in {0.5/1,2.5/2,4.5/3,7/4} {
    \node[figlabel] at (\x,3.87) {\(T_{\name}\)};
  }
  \foreach \x in {1.5,3.5,5.5} {
    \draw[black!45,densely dashed] (\x,2.53) -- (\x,3.45);
  }
  \node[figlabel] at (0,4.55) {prefix};
  \node[figlabel] at (3.5,4.55) {complete factors};
  \node[figlabel] at (8,4.55) {suffix};
  \node[figlabel,anchor=east] at (-0.6,0) {\(B=\)};
  \foreach \i/\letter in {0/c,1/c,2/b,3/b,4/a,5/b} {
    \node[figcell,fill=figInsert] at (\i,0) {\(\mathtt{\letter}\)};
  }
\foreach \source/\target in {0/0,3/1,2/2,5/3,4/4,8/5} {
    \draw[-{Stealth[length=1.8mm]},semithick,black!65]
      (\source,2.53) -- (\target,0.46);
  }
  \node[figcell,fill=figStatic,dashed] at (7.4,0) {\(\mathtt{d}\)};
  \node[figlabel] at (10,0) {carry-over~(not in \(B\))};
\end{tikzpicture}
\caption{A construction batch may cross Lyndon factors.  Here
\(T=\mathtt c\mathtt d^6\cdot\mathtt{bc}\cdot\mathtt{ab}\cdot\mathtt{aab}\),
where the single \(\mathtt d^6\) box abbreviates six consecutive
\(\mathtt d\)'s inserted in the preceding batch (also of length six);
arrows map the highlighted source letters to their positions in the
insertion-order batch, reversing their order within each factor.
The optional carry-over letter \(\mathtt d\), inserted in the preceding batch,
is charged separately to \(\Gamma_B\), accounting for the \(+1\) in
\(\sigma_\tau\le3\tilde{\sigma}_\tau+1\).}
\label{fig:batch-intervals}
\end{figure}

\Cref{fig:batch-intervals} illustrates the three source intervals.

Thus, passing from length-\(\tau\) substrings to construction batches
affects the space bound by at most a constant factor.

\begin{theorem} \label{thm:bbwt:basicTradeoff}
    Given a text \(T[1..n]\) and an integer \(\tau\) with \(\tau \in [2..n]\), 
    the BBWT of \(T\) can be constructed in
\( O\!\left(\frac{n^2}{\tau}\log \tau+n\sigma_\tau\right) \)
time using
\( O(\tau\sigma_\tau) \)
additional space, apart from the working array \(X\), where \(\sigma_\tau\le\min(\sigma,\tau+1)\).
The BBWT can be inverted within the same time and space bounds when replacing $\sigma_\tau$ with the global alphabet size \(\sigma\).
In particular, if \(\tau\sigma_\tau=O(n)\) for construction (respectively, \(\tau\sigma=O(n)\) for inversion),
the time simplifies to \(O((n^2/\tau)\log\tau)\).
\end{theorem}

\begin{proof}[Proof idea]
Duval's algorithm finds the Lyndon factors from left to right.
Before each batch, reverse newly discovered factors and prepare the next at most \(\tau\) letters in insertion order, 
immediately after the static transform in \(X\).  
A factor crossing a batch boundary is resumed using constant-size state.  
The batch alphabet is therefore known before the rank samples are built.
We mark factor starts, where insertion is at position~1; 
all other insertions follow \cref{eq:insertionrule}.
In total there are \(n\) insertions and \(O(n)\) LF steps.

The BBS (\cref{sec:batch-breakpoint-structure}) combines sampled ranks on the static transform with at most \(\tau\) pending breakpoints.
Its queries agree with those on the materialized transform.
A batch of static length \(z\) takes
\(O((z+\tau)\log\tau+\tau\sigma_\tau)\) time and
\(O(\tau\sigma_\tau)\) space, including counter initialization.
Over \(O(n/\tau)\) batches the total time is
\(
		O\!\left(\frac{n^2}{\tau}\log\tau+n\sigma_\tau\right)
\)
time.  
Duval's scans and factor reversals add \(O(n)\) total time.

Inversion extracts factors using select queries and a virtual delimiter.
Deleted positions and decoded letters are buffered until each flush;
the analogous analysis uses \(\sigma\) instead of \(\sigma_\tau\).
\end{proof}

\paragraph{The insertion rule.}
Let \(p\) be the position of the previously inserted letter of the
current Lyndon factor.  Insert the next letter \(\mathtt{c}\) at
\begin{equation}\label{eq:insertionrule}
    p' =
    \begin{cases}
        1 & \text{if \(\mathtt{c}\) is the first consumed letter of a new Lyndon factor,}\\
        \LF(p)+1 & \text{otherwise,}
    \end{cases}
\end{equation}
by inserting \(\mathtt{c}\) before the old \(L[p']\), then set \(p\gets p'\).
The LF step uses the transform \emph{before} insertion and inclusive rank.
The \(+1\) is essential: omitting it transforms \(\mathtt{ab}\) into
\(\mathtt{ab}\), not \(\BBWT(\mathtt{ab})=\mathtt{ba}\).

\begin{example}
Let \(T=\mathtt{abaab}\).
Its factors are \(T_1=\mathtt{ab}\), \(T_2=\mathtt{aab}\);
their reversals \(\mathtt{ba}\,\mathtt{baa}\) give the insertion order.
A bullet marks the first letter of a (reversed) factor we process.

\begin{center}
\begin{tabular}{clcll}
\toprule
step & consumed letter & new factor & insertion position \(p'\) & \(L\) after the insertion \\
\midrule
1 & \(\mathtt{b}=T_1[2]\) & \(\bullet\) & \(1\)                       & \(\mathtt{b}\) \\
2 & \(\mathtt{a}=T_1[1]\) &             & \(\LF(1)+1=1+1=2\)          & \(\mathtt{ba}\) \\
3 & \(\mathtt{b}=T_2[3]\) & \(\bullet\) & \(1\)                       & \(\mathtt{bba}\) \\
4 & \(\mathtt{a}=T_2[2]\) &             & \(\LF(1)+1=2+1=3\)          & \(\mathtt{bbaa}\) \\
5 & \(\mathtt{a}=T_2[1]\) &             & \(\LF(3)+1=1+1=2\)          & \(\mathtt{babaa}\) \\
\bottomrule
\end{tabular}
\end{center}

\noindent
Indeed, the conjugates in \(\omega\)-order are
\(\mathtt{aab},\mathtt{aba},\mathtt{ab},\mathtt{baa},\mathtt{ba}\),
whose cyclic predecessors give \(\BBWT(T)=\mathtt{babaa}\).
\end{example}

\paragraph{Remark on the intermediate transforms.}
Intermediate transforms need not equal the BBWT of consumed letters:
after step~3, \(L=\mathtt{bba}\), whereas the extended BWT of
\(\{\mathtt{ab},\mathtt{b}\}\) is \(\mathtt{bab}\).
So our algorithm does not incrementally build a valid BBWT like an online algorithm, but it obeys the following invariant.

Let \(T=T_1\cdots T_\ell\) be the Lyndon factorization and fix \(x\in[1..\ell]\).
Assume that, before \(T_x\) is processed, the current transform is \(\BBWT(T_1\cdots T_{x-1})\), with the empty transform for \(x=1\), and let \(m=|T_x|\).
Insert \(T_x[m]\) at position~1 and set \(p\gets1\).
For \(j=m-1,m-2,\ldots,1\), evaluate \(\LF\) on the current transform, set \(p\gets\LF(p)+1\), and insert \(T_x[j]\) at position~\(p\).
Throughout this loop, the following invariant holds due to \cite[Lemma~2]{koppl20inplace}.
    \begin{equation}
            \label{inv:insertion}\tag{I1}
	\begin{aligned}
&	        \text{\(p\) is the position of the letter of \(T_x\) most recently inserted, and} \\
&	    \text{after the loop the current transform is \(\BBWT(T_1\cdots T_x)\).}
	\end{aligned}
    \end{equation}
Consequently, after all Lyndon factors have been processed, the current transform is \(\BBWT(T)\).

\section{Letter-Piece Tables}
\label{sec:lpt}

Construction batches contain only insertions.  
However, writing updates of $L$ directly into $Z$ is costly since an insertion at position $i$ needs to move all letters at positions $i..|L|$ to the right.
Instead, we represent $L$ logically: 
we keep $Z$ static and bookkeep the insertions in such a way that we can answer queries on the logical transform \(L\) without materializing it ---
we materialize $L$ after at most \(\tau\) insertions.
This bookkeeping can be formalized in terms of a \emph{letter-piece table} (LPT) data structure.

\problembox{\textsc{Letter-Piece Table} (LPT)\\
An LPT represents a text $L$ implicitly and provides update operations to insert letters at specified positions.
The text $L$ is initially given explicitly as a static string $Z[1..z]$, which is kept unchanged by the LPT until materialization is invoked\@.
The LPT provides the following operations:
\begin{itemize}
    \item $L.\myFunction{insert}(i,\mathtt{c})$: inserts $\mathtt{c}$ before $L[i]$, i.e., sets $L[i+1..] \gets L[i..]$ and $L[i] \gets \mathtt{c}$ for $i \in [1..|L|+1]$,
    \item $L.\myFunction{access}(i)$: returns $L[i]$,
    \item $L.\myFunction{materialize}$: materializes the represented $L$ and replaces $Z$ with $L$
\end{itemize}
}

We implement an LPT with a two-level balanced binary search tree (BST) structure.
First, a \emph{breakpoint} is a triple
\(
    b=\langle \mathtt{c}_b,j_b,\rho_b\rangle,
\)
where \(\mathtt{c}_b\) is the inserted letter, \(j_b\in[1..z+1]\) is its static anchor 
(before \(Z[j_b]\), or after \(Z\) for \(j_b=z+1\)),
and \(\rho_b\) is a handle that specifies its order among breakpoints with that anchor.

An outer BST \(\mathcal B\), 
keyed by static coordinates, 
stores the nonempty buckets \(\mathcal B_j\) of breakpoints anchored at \(j\),
plus a possibly empty sentinel bucket at \(z+1\).

Each bucket is an order-statistic BST whose inorder traversal is the local emission order \(\prec_{\mathcal B_j}\).
The handle \(\rho_b\) identifies a node or adjacent boundary, 
so insertion at a local boundary takes \(O(\log|\mathcal B_j|)\) time.

Materialization emits each gap followed by its bucket:
Let \(j_1<\cdots<j_s=z+1\) be the bucket coordinates and set \(j_0=1\).  
Before bucket \(\mathcal B_{j_t}\) lies the static gap
\(
    Z[j_{t-1}..j_t-1],
\)
possibly empty.  

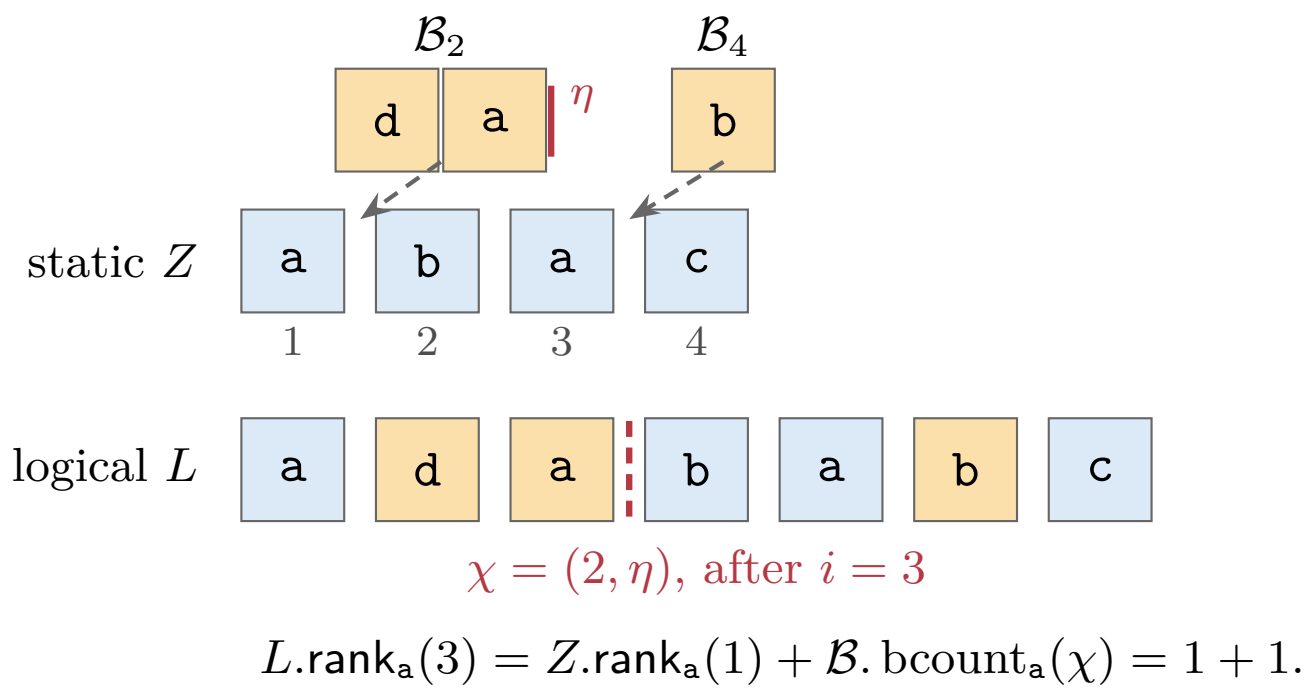
\begin{figure}[t]
\centering
\begin{tikzpicture}[x=0.89cm,y=0.69cm]
  \node[figlabel,anchor=east] at (-0.55,3) {static \(Z\)};
  \foreach \i/\letter in {0/a,1/b,2/a,3/c} {
    \node[figcell,fill=figStatic] at (\i,3) {\(\mathtt{\letter}\)};
    \node[font=\scriptsize,text=black!70] at (\i,2.23) {\(\the\numexpr\i+1\relax\)};
  }
  \node[figlabel] at (1.1,5.15) {\(\mathcal B_2\)};
  \node[figcell,fill=figInsert] at (0.7,4.35) {\(\mathtt d\)};
  \node[figcell,fill=figInsert] at (1.5,4.35) {\(\mathtt a\)};
  \draw[figarrow,black!60,densely dashed] (1.1,3.95) -- (0.5,3.4);
  \draw[figMarker,very thick] (1.92,4.0) -- (1.92,4.68);
  \node[font=\scriptsize,text=figMarker] at (2.15,4.55) {\(\eta\)};
  \node[figlabel] at (3.2,5.15) {\(\mathcal B_4\)};
  \node[figcell,fill=figInsert] at (3.2,4.35) {\(\mathtt b\)};
  \draw[figarrow,black!60,densely dashed] (3.2,3.95) -- (2.5,3.4);
  \node[figlabel,anchor=east] at (-0.55,1) {logical \(L\)};
  \foreach \i/\letter/\shade in {0/a/figStatic,1/d/figInsert,2/a/figInsert,3/b/figStatic,4/a/figStatic,5/b/figInsert,6/c/figStatic} {
    \node[figcell,fill=\shade] at (\i,1) {\(\mathtt{\letter}\)};
  }
  \draw[figMarker,very thick,densely dashed] (2.5,0.55) -- (2.5,1.55);
  \node[figlabel,text=figMarker] at (3.0,0.03)
        {\(\chi=(2,\eta)\), after \(i=3\)};
  \node[figlabel,anchor=west] at (-0.4,-0.83)
        {\(\displaystyle L.\fnRank_{\mathtt a}(3)
          = Z.\fnRank_{\mathtt a}(1)
          +\mathcal B.\operatorname{bcount}_{\mathtt a}(\chi)
          =1+1.\)};
\end{tikzpicture}
\caption{A letter-piece table with static letters (blue) and pending
breakpoints (orange).  The buckets \(\mathcal B_2\) and \(\mathcal B_4\)
are anchored before the corresponding static letters; a local boundary
\(\eta\) determines the logical cut \(\chi\).  The prefix
\(L[1..3]=\mathtt{ada}\), to the left of the dashed cut, contains
one static and one inserted \(\mathtt a\).}
\label{fig:lpt-cut}
\end{figure}

To locate positions in \(L\), define a \emph{cut}
\(
    \chi=(j,\eta),
\)
where \(j\in[1..z+1]\) and \(\eta\) is a boundary of
\(\mathcal B_j\).
The cut denotes the prefix consisting of \(Z[1..j-1]\), all breakpoints
anchored before \(j\), and those of \(\mathcal B_j\) preceding \(\eta\).
\Cref{fig:lpt-cut} illustrates the representation and a cut.

\paragraph{Materialization.}
If the breakpoints in global order are
\( b_1<b_2<\cdots<b_m, \)
with $b_t=\langle \mathtt{c}_t,j_t,\rho_t\rangle$,
then 
\(
    L = Z[1..j_1-1],\ \mathtt{c}_1,\ Z[j_1..j_2-1],\ \mathtt{c}_2,\ \ldots,\ \mathtt{c}_m,\ Z[j_m..z],
\)
where equal anchors give empty intervening gaps.

To write \(L\) into \(X[1..z+m]\) without overwriting unread
letters of \(Z\), traverse the buckets and gaps \emph{right to left}.
Every static letter originally at \(i\) moves to an index at least
\(i\), so its destination has already been vacated.
The sweep visits \(z+m\) letters and takes \(O(z+\tau)\) time.
For inversion, we use a deletion-only variant of the LPT (replacing the operation insert with delete), whose
materialization instead scans the static string left to right (\cref{sec:inversion}).

\paragraph{Access and insertion.}
Let \(\myFunction{accessCut}(i)\) return the cut after the first
\(i-1\) letters of \(L\).  Each outer node stores its contribution
\(g_j+|\mathcal B_j|\), where \(g_j\) is the preceding static-gap
length, as well as subtree-length sums.
Creating a bucket inside a gap splits that gap and updates its
successor's length.

Weighted selection in the outer tree finds the relevant gap or bucket
in \(O(\log|\mathcal B|)\) time.  Within a gap, arithmetic gives the
static coordinate; within a bucket, local selection finds \(\eta\).
Both \(\myFunction{accessCut}(i)\) and \(\myFunction{access}(i)\)
therefore take \(O(\log\tau)\) time.

To insert \(\mathtt{c}\), find \(\chi=\myFunction{accessCut}(i)=(j,\eta)\),
create bucket \(\mathcal B_j\) if necessary, insert a breakpoint at
\(\eta\), and update the length summaries.

\begin{lemma}\label{lem:lpt}
    There is an LPT that supports \(\myFunction{access}(i)\) and \(\myFunction{insert}(i,\mathtt{c})\) in \(O(\log\tau)\) time, 
    and that can materialize $L$ in \(O(z+\tau)\) time, where $z=|Z|$.
    It uses \(O(\tau)\) space on top of the static string \(Z\).
\end{lemma}

\section{The Batch-Breakpoint Structure (BBS)}
\label{sec:batch-breakpoint-structure}

The BBS augments an LPT with rank and cumulative-count queries.
We first describe batch preparation, then the additional summaries.

\subsection{Preparing a construction batch}
\label{subsec:prepare-construction-batch}

Before building rank samples on the static transform of length \(z\),
prepare \(r\le\tau\) source letters so that
\(
	X[z+1..z+r]
\)
contains the next \(r\) letters in BBWT insertion order.

Run Duval's algorithm when the next factor is needed, and reverse each detected factor \(X[f..e]\) in place.  
If the batch ends within it, retain the next unread coordinate \(u\) and factor end \(e\).
A constant-size state \(\mathcal Q\) records either Duval's next start position or \((u,e)\), 
a factor-start flag, and the boundary LF letter when needed.  
Thus, a factor is detected and reversed only once.

We record the factor starts in a set of positions \(\mathcal I\subseteq[1..r]\).
Given we insert a letter \(\mathtt{c}\) from a position in $\mathcal I$ into $L$,
put it at position \(1\); 
otherwise insert at \(\LF(p)+1\), as dictated by \cref{eq:insertionrule}.
The construction-batch alphabet is
\[
	\Gamma_B = \Gamma(X[z+1..z+r])
	\cup
	\{\text{the boundary LF letter, if present}\}.
\]
The optional extra letter belongs to the last insertion of the
previous batch, needed for the first LF step when a factor continues.
Thus, \(|\Gamma_B|\le\tau+1\); 
the preparation state, \(\mathcal I\), and the alphabet dictionary use \(O(\tau)\) space.

\subsection{Construction breakpoints as a bounded edit buffer}
\label{subsec:bbs-structure}

The BBS adds two operations to the LPT:

\begin{itemize}
    \item $L.\fnRank_{\mathtt{c}}(i)$ returns the number of occurrences of letter \(\mathtt{c}\) in the prefix \(L[1..i]\).
    \item $L.\fnCount(\mathtt{c})$ returns the number of occurrences of letters smaller than \(\mathtt{c}\) in \(L\).
\end{itemize}

We call this augmented LPT a \emph{batch-breakpoint structure} (BBS).

To answer rank queries,
we let each local tree \(\mathcal B_j\) store subtree frequencies for all \(\mathtt{c}\in\Gamma_B\).

For the ordinary BWT, the delimiter is carried as separate state,
as in \citet{crochemore15bwt}.

\myblock{Static rank information on \(Z\)}
Partition \(Z\) into \(O(\tau)\) segments of length \(O(z/\tau+1)\), 
storing per-letter prefix counts at their boundaries.  
Finally, to answer $Z.\fnRank\_{\mathtt c}(j)$, 
locate $\mathtt c$ in the batch alphabet $\Gamma_B$ in $O(\log \sigma_\tau)$ time,
retrieve its preceding prefix-count sample, 
and scan the remainder of the containing segment. 
This all takes $O(z/\tau+\log\sigma\_\tau+1)$ time.

Also store \(C_Z[\mathtt{c}]=|\{h:Z[h]<\mathtt{c}\}|\) for \(\mathtt{c}\in\Gamma_B\).
For sorted \(\Gamma_B=\{\mathtt{c}_1<\cdots<\mathtt{c}_g\}\), scan \(Z\) once:
for each \(\mathtt{x}\in Z\), increment the counter at the smallest
\(a\) with \(\mathtt{x}<\mathtt{c}_a\), if any.  Prefix sums of these counters give
all \(C_Z[\mathtt{c}_a]\).

These arrays occupy \(O(\tau\sigma_\tau)\) space and are rebuilt
once per batch.

\myblock{Breakpoint rank}
Write \(b\prec\chi\) when breakpoint \(b\) precedes cut
\(\chi=(j,\eta)\); equivalently,
\[
    b = (\mathtt{c}_b,j_b,\rho_b) \prec \chi \Longleftrightarrow j_b<j \text{ or } \bigl(j_b=j \text{ and } \rho_b \prec_{\mathcal{B}_j} \eta \bigr). 
\]

Define the breakpoint prefix count
\[
    \mathcal{B}.\operatorname{bcount}_{\mathtt{c}}(\chi)
:=
|\{\,b=(\mathtt{c}_b,j_b,\rho_b) \in \mathcal B :
\mathtt{c}_b=\mathtt{c} \text{ and } b\prec\chi\,\}|.
\]
For a cut \(\chi=(j,\eta)\), this value decomposes as
\[
    \mathcal B.\operatorname{bcount}_{\mathtt{c}}(\chi)
    =|\{\,b\in\mathcal B:j_b<j,\ \mathtt{c}_b=\mathtt{c}\,\}|
    +|\{\,b\in\mathcal B_j:\mathtt{c}_b=\mathtt{c},\ \rho_b\prec_{\mathcal B_j}\eta\,\}|.
\]
Store per-letter subtree counts in the outer tree~$\mathcal B$, 
with each bucket's total as its node weight.  
An outer prefix-count query gives the first term in \(O(\log\tau)\) time; a local query on $\mathcal B_j$ gives the second.
These summaries use \(O(\tau\sigma_\tau)\) space.
Consequently,
\[
    L.\fnRank_{\mathtt{c}}(i)
    =
    Z.\fnRank_{\mathtt{c}}(j-1)+ \mathcal{B}.\operatorname{bcount}_{\mathtt{c}}(\chi),
\]
where \(\chi=(j,\eta)=\myFunction{accessCut}(i+1)\).
Thus rank takes \(O(z/\tau+\log\tau+1)\) time.

\myblock{Count queries}
The cumulative count used by LF is
\[
    L.\fnCount(\mathtt{c})
    :=
    C_Z[\mathtt{c}]
    +
    |\{\,\langle \mathtt{d},j,\rho\rangle \in \mathcal B : \mathtt{d}<\mathtt{c}\,\}|.
\]
Maintain the second term in a letter-keyed BST \(\mathcal D\), storing
the multiplicity of each pending letter and subtree multiplicity sums.
It has at most \(\min(\sigma_\tau,\tau)\) keys, so prefix queries
and updates take \(O(\log\tau)\) time and \(O(\sigma_\tau)\) space.
This avoids an \(O(\sigma_\tau)\) scan per update.
Finally,
\[
    \LF(i) = L.\fnCount(\mathtt{c}) + L.\fnRank_{\mathtt{c}}(i),
    \text{where~} \mathtt{c} = L.\myFunction{access}(i).
\]

\section{Inversion}\label{sec:inversion}
Following \citet[Section~5.3]{koppl20inplace}, 
we invert the BBWT by extracting \(T_\ell,T_{\ell-1},\ldots,T_1\).
Temporarily appending the delimiter \(\$\) to the current last factor allows us to perform the destructive inversion of \citet{crochemore15bwt}.
We represent the delimiter $\$$ either at a static coordinate or \emph{virtually}, 
by storing only its position as a cut $\mu$ between surviving letters, 
without occupying a cell of $Z$.
We decode each factor left to right, while writing factors right to left as space is released by batch flushes.

Proceed as follows:
Store the temporary delimiter's position as \(\mu\), with \(\mu=\bot\) when absent.  
At a factor start, insert it logically at position~2, as justified by \citet[Lemma~2]{koppl20inplace}.
Each FL step identifies the next output letter; output it, remove the old delimiter, and replace that letter by \(\$\).
An old delimiter at a static coordinate becomes a deletion; 
a virtual delimiter is simply removed.  
When FL returns to \(\$\), remove the marker and start the preceding factor.

The marker \(\mu\) is either a virtual cut or a static coordinate
temporarily replaced by \(\$\).
At a flush, copy only surviving ordinary letters and retain the
marker as a virtual cut, so \(X\) needs no extra delimiter cell.

The inversion algorithm uses a \emph{deletion-only LPT}:
an inverse batch records deleted ordinary-letter coordinates in a BST \(\mathcal E\), 
rather than insertion breakpoints, and needs no local bucket trees.  
The temporary delimiter is handled separately by \(\mu\).
Hence, \(\mathcal E\) is a BST that bookkeeps the deleted text positions.

For \(\FL(i)\), 
first find the letter \(\mathtt{c}=F[i]\) from the cumulative counts \(C_L\), then compute
\(
    \FL(i)= L.\fnSelect_{\mathtt{c}}(i-C_L[\mathtt{c}]).
\)
Here \(L\) is \(Z\) with deletions \(\mathcal E\) and the marker \(\mu\).

Augment \(\mathcal E\) with subtree sizes and deleted-letter
frequencies over the global alphabet \(\Sigma\).
Let \(\mathcal E^\star\) contain both deleted coordinates and the
coordinate occupied by the marker, if any; a virtual marker contributes
no coordinate.  
The tree \(\mathcal E^\star \) supports
\[
    \operatorname{dcount}_{\mathtt{c}}(j)
    :=
    |\{\,h\in\mathcal E^\star : h\le j \text{ and } Z[h]=\mathtt{c}\,\}|
\]
in \(O(\log\tau)\) time.

If \(j_i\) is the static boundary (in $Z$) after the first \(i\) logical letters (in $L$), 
excluding the marker when counting static coordinates, then
\(L.\fnRank_{\mathtt{c}}(i)=Z.\fnRank_{\mathtt{c}}(j_i)-\operatorname{dcount}_{\mathtt{c}}(j_i)\)
for \(\mathtt{c}\in\Sigma\).  For select, find
\( h = \min\{h' : Z.\fnRank_{\mathtt{c}}(h')-\operatorname{dcount}_{\mathtt{c}}(h')\ge q\} \),
and convert \(h\) to its logical position using the number of
excluded coordinates and the marker position.

Partition \(Z\) into \(O(\tau)\) blocks of length
\(\lceil z/\tau\rceil\).  For every \(\mathtt{c}\in\Sigma\), maintain
dynamic prefix sums of surviving occurrences per block.
Rank and select use these sums and scan at most one block; during a
scan, traverse the sorted deleted coordinates without repeated tree
searches.  Each deletion updates one block count.
Both queries take
\(
    O(\log\tau+|Z|/\tau)
\)
time and the structure uses \(O(\tau\sigma)\) space.

As in construction, a letter-keyed prefix-sum tree with at most \(\min(\sigma,\tau)\) keys counts excluded letters below \(\mathtt{c}\) and supports updates in \(O(\log\tau)\) time.

The updated cumulative counts are
\(
C_L[\mathtt{c}]
=
C_Z[\mathtt{c}]
-
|\{h\in\mathcal E^\star : Z[h]<\mathtt{c}\}|
+
\Delta_\mu(\mathtt{c}),
\)
where
\[
\Delta_\mu(\mathtt{c})=
\begin{cases}
1 & \text{if } \mu\neq\bot \text{ and } \mathtt{c}\in\Sigma,\\
0 & \text{otherwise.}
\end{cases}
\]

\begin{figure}[t]
\centering
\begin{tikzpicture}[x=0.83cm,y=0.80cm]
  \node[figlabel,anchor=east] at (-0.6,3.2) {static \(Z\)};
  \foreach \i/\letter/\shade in {0/a/figStatic,1/b/figExcluded,2/c/figStatic,3/a/figStatic,4/b/figStatic,5/c/figExcluded,6/a/figStatic} {
    \node[figcell,fill=\shade] at (\i,3.2) {\(\mathtt{\letter}\)};
    \node[font=\scriptsize,text=black!65] at (\i,2.43)
          {\(\the\numexpr\i+1\relax\)};
  }
  \foreach \i in {1,5} {
    \draw[figMarker,thick] (\i-0.27,2.93) -- (\i+0.27,3.47);
    \draw[figMarker,thick] (\i-0.27,3.47) -- (\i+0.27,2.93);
  }
  \draw[figMarker,very thick] (3.5,2.9) -- (3.5,3.55);
  \node[figlabel,text=figMarker] at (3.65,3.95) {\(\mu\)};
  \node[figlabel,anchor=west] at (7.0,3.2)
        {\(\mathcal E^\star=\{2,6\}\)};

  \node[figlabel,anchor=east] at (-0.6,1.55) {flushed \(Z'\)};
  \foreach \i/\letter in {0/a,1/c,2/a,3/b,4/a} {
    \node[figcell,fill=figStatic] at (\i,1.55) {\(\mathtt{\letter}\)};
  }
  \draw[figMarker,very thick] (2.5,1.2) -- (2.5,1.9);
  \node[figlabel,text=figMarker,anchor=west] at (5,1.55)
        {same virtual cut};

  \node[figlabel,anchor=east] at (-0.6,0) {logical \(L\)};
  \foreach \i/\letter in {0/a,1/c,2/a,4/b,5/a} {
    \node[figcell,fill=figStatic] at (\i,0) {\(\mathtt{\letter}\)};
  }
  \node[figcell,fill=figInsert,draw=figMarker,very thick,
        text=figMarker] at (3,0) {\(\$\)};
  \draw[figarrow,black!60] (6.5,2.65) -- (6.5,1.9)
        node[midway,right,font=\scriptsize] {flush};
\end{tikzpicture}
\caption{A deletion-only LPT with a virtual delimiter.
The excluded static coordinates (gray, crossed out) are omitted,
and the delimiter is inserted at the cut \(\mu\).
Flushing compacts the surviving letters into \(Z'\) and retains
the cut, so both representations yield
\(L=\mathtt{aca}\$\mathtt{ba}\), as required by invariant~\ref{inv:inverse-flush}.}
\label{fig:inverse-delimiter}
\end{figure}
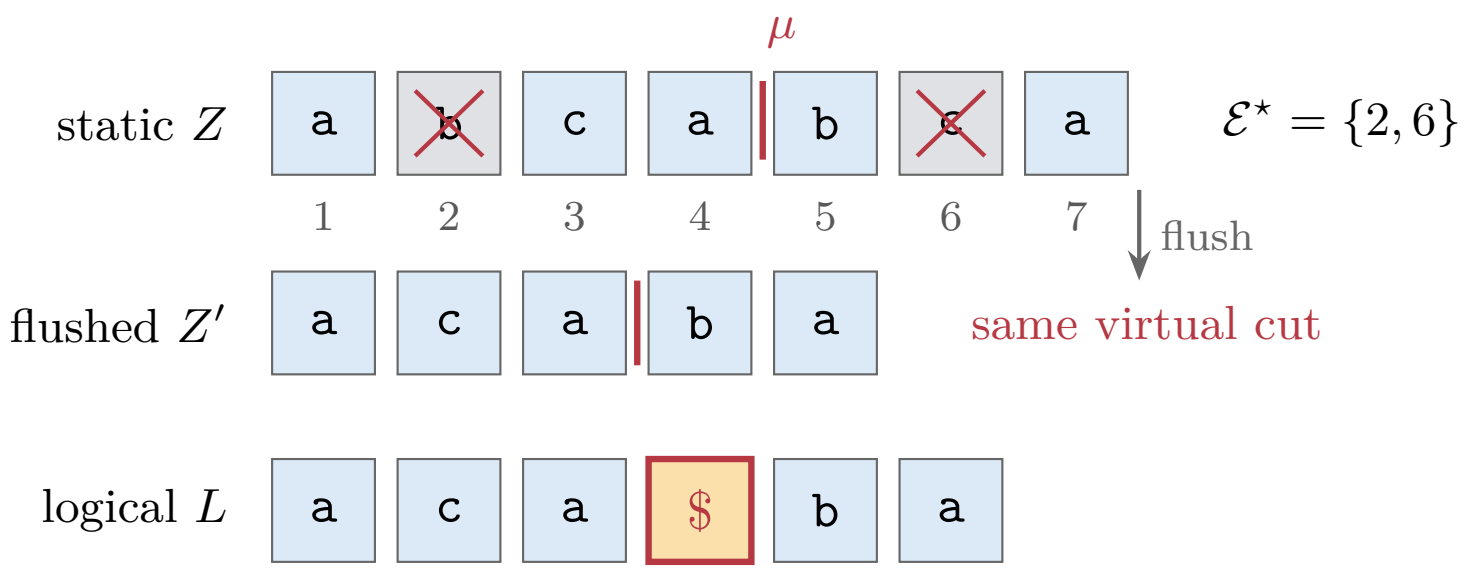

Materialization scans \(Z\), 
skips the deleted positions bookkept in \(\mathcal E^\star\), 
and updates the position \(\mu\) of the virtual delimiter in the compacted string, in \(O(z+\tau)\) time.
By doing so, the logical transform is unchanged, as depicted in \Cref{fig:inverse-delimiter}.

\paragraph{Writing the decoded factors back.}
Factors are decoded left to right but occupy the output from right to
left.  Their lengths are initially unknown, so write each factor
temporarily in reverse order, growing leftward.  Let \(\beta\) be the
next free position to the left of all output written so far.
Buffer the at most \(\tau\) letters decoded in a batch, together with
their factor boundaries; do not overwrite the static transform before
the flush.

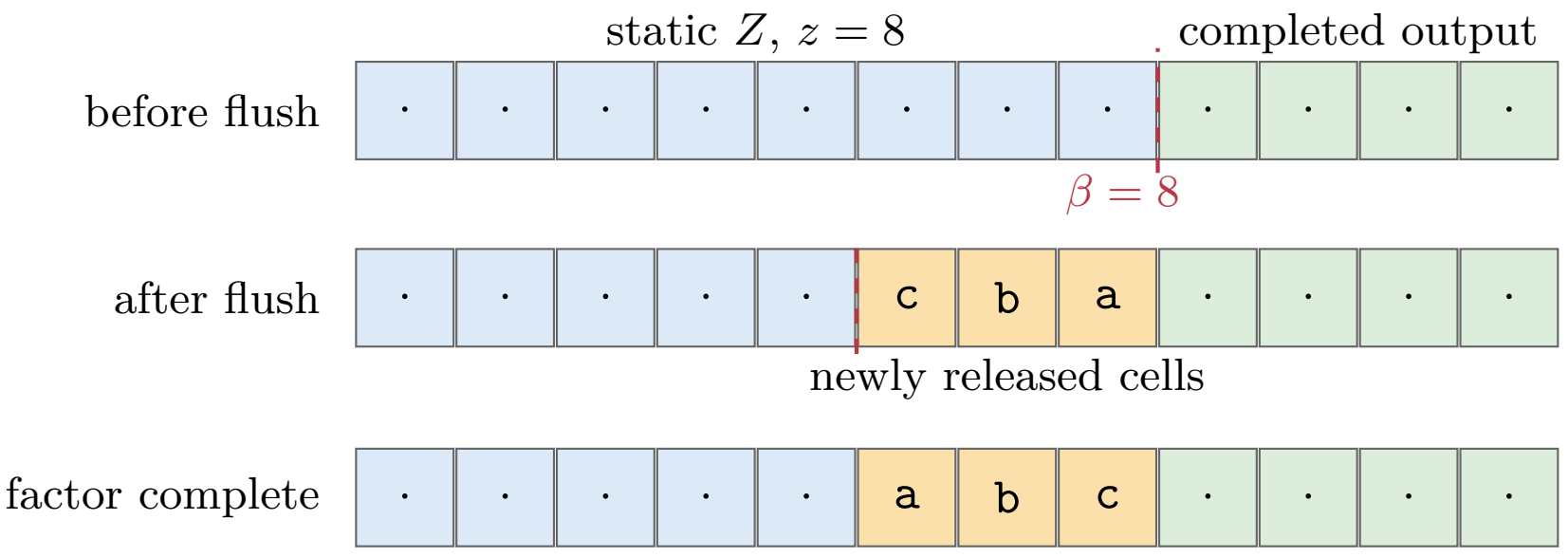
\begin{figure}[t]
\centering
\begin{tikzpicture}[x=0.70cm,y=0.87cm]
  \node[figlabel,anchor=east] at (-0.65,3) {before flush};
  \foreach \i in {0,...,7} {
    \node[figcell,fill=figStatic] at (\i,3) {\(\cdot\)};
  }
  \foreach \i in {8,...,11} {
    \node[figcell,fill=figOutput] at (\i,3) {\(\cdot\)};
  }
  \node[figlabel] at (3.5,3.62) {static \(Z\), \(z=8\)};
  \node[figlabel] at (9.5,3.62) {completed output};
  \draw[figMarker,thick,dashed] (7.5,2.5) -- (7.5,3.5);
  \node[figlabel,text=figMarker] at (7.15,2.33) {\(\beta=8\)};

  \node[figlabel,anchor=east] at (-0.65,1.5) {after flush};
  \foreach \i in {0,...,4} {
    \node[figcell,fill=figStatic] at (\i,1.5) {\(\cdot\)};
  }
  \foreach \i/\letter in {5/c,6/b,7/a} {
    \node[figcell,fill=figInsert] at (\i,1.5)
          {\(\mathtt{\letter}\)};
  }
  \foreach \i in {8,...,11} {
    \node[figcell,fill=figOutput] at (\i,1.5) {\(\cdot\)};
  }
  \draw[figMarker,thick,dashed] (4.5,1.05) -- (4.5,1.95);
  \node[figlabel] at (6,0.85) {newly released cells};

  \node[figlabel,anchor=east] at (-0.65,-0.1) {factor complete};
  \foreach \i in {0,...,4} {
    \node[figcell,fill=figStatic] at (\i,-0.1) {\(\cdot\)};
  }
  \foreach \i/\letter in {5/a,6/b,7/c} {
    \node[figcell,fill=figInsert] at (\i,-0.1)
          {\(\mathtt{\letter}\)};
  }
  \foreach \i in {8,...,11} {
    \node[figcell,fill=figOutput] at (\i,-0.1) {\(\cdot\)};
  }
  \draw[figarrow,black!70] (11.95,2.8) -- (11.95,1.75);
  \draw[figarrow,black!70] (11.95,1.25) -- (11.95,0.18);
\end{tikzpicture}
\caption{In-place output placement for a batch decoding
\(\mathtt a,\mathtt b,\mathtt c\) (\(z=8\), \(r=3\)).
Compacting the surviving transform to \(X[1..z-r]\) frees
\(X[6..8]\); writing leftward from \(\beta=z\) stores
\(\mathtt{cba}\) there.  After the factor is finished, reversing
its output interval restores \(\mathtt{abc}\).
Previously completed factors (green) remain untouched.}
\label{fig:inverse-writeback}
\end{figure}

If the batch decodes \(r\) letters from a static transform of length
\(z\), first compact the surviving ordinary transform into
\(X[1..z-r]\).  Then write the buffered letters in decoding order to
\(X[\beta],X[\beta-1],\ldots,X[\beta-r+1]\), restoring original
letters if necessary.  Since \(\beta=z\) at the start of the batch,
these are exactly the freed cells.  Continue across batches if a
factor is unfinished.  Once a factor is complete, reverse its entire
interval, including letters written in earlier batches; reversals
occur only after the relevant flush.  Factor boundaries use
\(O(\tau)\) space per batch and all reversals take \(O(n)\) time.
No unflushed static letter is overwritten.
\Cref{fig:inverse-writeback} illustrates the reuse of the freed cells.

An inverse batch performs at most \(\tau\) destructive FL steps,
records deletions in \(\mathcal E\), buffers decoded letters, and
maintains \(\mu\).  Rank/select costs \(O(|Z|/\tau+\log\tau)\) per
step, with \(O(\tau\sigma)\) auxiliary space.

\paragraph{Inversion invariants.}
Let \(T=T_1\cdots T_\ell\) be the Lyndon factorization of $T$, and define \(P_x := T_1\cdots T_{x-1}\).
At the start of extracting \(T_x\), 
the remaining BBWT is \(\BBWT(P_xT_x)\).
After inserting the unique delimiter at position~\(2\), 
let \(j\in[0..|T_x|]\) be the number of extracted letters of \(T_x\).
The \emph{inverse-factor invariant} is
\begin{equation}\label{inv:inverse-factor}\tag{I2}
    L=\BBWT\bigl(P_x\,\$\,T_x[j+1..|T_x|]\bigr).
\end{equation}
When \(j=|T_x|\), removing the delimiter leaves \(\BBWT(P_x)\).

For the \emph{batch-boundary invariant}, let the logical
transform \(L\) be represented by a static string \(Z\), excluded
static coordinates \(\mathcal E^\star\), and the delimiter state
\(\mu\), possibly \(\bot\).
A flush copies the surviving ordinary entries of \(Z\) in their
original order and carries the delimiter's cut when present.
The invariant asserts that the logical transform is preserved:
\begin{equation}\label{inv:inverse-flush}\tag{I3}
    L_{\mathrm{after\ flush}}=L_{\mathrm{before\ flush}}.
\end{equation}

\begin{lemma}\label{lem:inverse-invariants}
Invariants~\ref{inv:inverse-factor} and~\ref{inv:inverse-flush} hold.
\end{lemma}

\begin{proof}
\emph{Invariant~\ref{inv:inverse-factor}.}
The first entry of the BBWT of a nonempty text is the last
letter of its last Lyndon factor
\cite[Lemma~2]{koppl20inplace}.
As shown in \cite[Section~5.3]{koppl20inplace},
inserting \(\$\) at position~\(2\) extends the cycle of the
last Lyndon factor \(T_x\) to that of the Lyndon word
\(\$T_x\).  Every nonempty suffix \(V\) of \(T_x\)
has \(\$V\) Lyndon, because \(\$\) is unique and smaller
than every ordinary letter.
The destructive forward step of
\citet[Fig.~3]{crochemore15bwt}, applied to the
extended last factor, outputs its first ordinary letter,
removes the old delimiter entry, and replaces the
entry of that letter by the delimiter.
It thereby changes the last factor from
\(\$T_x[j+1..]\) to \(\$T_x[j+2..]\), leaving the
previous Lyndon factors unchanged.
Induction on \(j\) proves Invariant~\ref{inv:inverse-factor}.
After the last ordinary letter is extracted, the
factor is the singleton \(\$\); removing it restores
\(\BBWT(P_x)\).

\emph{Invariant~\ref{inv:inverse-flush}.}
Let \(U=\operatorname{delete}_{\mathcal E^\star}(Z)\).
The logical transform is \(U\) with \(\$\), if present,
inserted at the cut represented by \(\mu\).
A flush writes \(U\) as the new static string and
preserves this cut, hence the logical transform and all
its rank, select, and \(\FL\) values.
\end{proof}

\section{Space and time}
Construction stores \(O(\tau)\) sampled counters per batch letter and \(O(\tau)\) breakpoints with per-letter summaries.
The arrays \(C_Z\), pending frequencies, 
and \(\mathcal D\) add \(O(\sigma_\tau)\) words.  
Thus, its auxiliary space is \(O(\tau\sigma_\tau)\); 
inversion uses \(O(\tau\sigma)\) space.
Both work in the original array \(X\), retaining edits until a flush with an LPT\@.

With BSTs, building rank samples on a static transform of length \(z\) takes \(O(z(1+\log\sigma_\tau))\subseteq O(z(1+\log\tau))\) time: 
each scanned letter requires one batch-alphabet predecessor query.
Each LF, rank-at-cut, access-cut, or insertion costs \(O(z/\tau+\log\tau)\) time; 
there are at most \(\tau\) updates per batch.
Inversion has the same batch bound, and materialization costs \(O(z+\tau)\) time.
Hence, each batch takes \(O((z+\tau)\log\tau+\tau\gamma)\) time, 
where \(\gamma=\sigma_\tau\) for construction and \(\gamma=\sigma\) for inversion.  
Summing over \(O(n/\tau)\) batches gives
\(
		O\!\left(\frac{n^2}{\tau}\log\tau+n\gamma\right)
\)
time, or \(O((n^2/\tau)\log\tau)\) time if \(\tau\gamma=O(n)\).

\section{B-Tree Acceleration of the Prefix-Sum Searches}
\label{sec:btree}
On a word RAM with \(w\in\Omega(\log n)\), high-degree B trees
with aggregate sums~\cite{i25spaceefficient} reduce the BBS search
time to \(O(\log\tau/\log\log\tau)\).
We stipulate that letters are ordered integer keys fitting in one word.

We handle three operations: static predecessor search in the batch
alphabet, weighted selection in \(L\), and prefix counts of pending
letters.  Selection must account for static gaps of arbitrary lengths
between breakpoints; these gaps are stored at leaves, with searchable
partial sums at internal nodes.

\problembox{\textsc{Small Searchable Partial Sums} (SSPS)\\
Fix a word-RAM word size \(w\) and a length \(\delta\ge2\).
Maintain $\delta$ nonnegative integers
\(a_1,\ldots,a_\delta\) 
whose total \(M:=\sum_{i=1}^\delta a_i\) fits in one \(w\)-bit word,
i.e., \(0\le M<2^w\).
The initial entries \(a_1,\ldots,a_\delta\) are given explicitly.
Support the following operations:
\begin{itemize}
    \item \(\operatorname{sum}(k)\), for \(k\in[0..\delta]\):
    return \(\sum_{i=1}^k a_i\), where \(\operatorname{sum}(0)=0\).
    \item \(\operatorname{search}(q)\), for \(q\in[1..M]\):
    return the smallest \(k\in[1..\delta]\) with
    \(\operatorname{sum}(k)\ge q\).\\
    Entries $a_i$ may be zero, so the smallest qualifying index is required.
    \item \(\operatorname{update}(k,\xi)\), for \(k\in[1..\delta]\)
    and \(\xi\in\{-1,+1\}\): set \(a_k\gets a_k+\xi\).\\
    Every update must preserve the conditions \(a_k\ge0\) and \(0\le M<2^w\).
\end{itemize}
}

\begin{lemma}[Small searchable partial sums]\label{lem:btree-micro}
Fix a sufficiently small constant \(\varepsilon>0\) and
\(2\le\delta=O(w^\varepsilon)\).
SSPS can be initialized in \(O(\delta)\) time using \(O(\delta)\) words,
and supports each operation in \(O(1)\) amortized time.
\end{lemma}
\begin{proof}
We adapt the small-array structure of \citet[Section~2.2]{patrascu04tight}. 
Write \(S[i]=\operatorname{sum}(i)\).

At initialization and after every \(\delta\) updates, 
we \emph{freeze} the prefix sums, similarly to what we did with $Z$ in LPT\@:
we store their current values \(S_0[i]\) and leave these copies unchanged until the next rebuild.
We partition the indices into \emph{runs}, 
that is, maximal contiguous intervals of indices whose successive frozen prefix sums differ by less than \(4\delta\).
Thus, a new run starts at \(i>1\) exactly when \(S_0[i]-S_0[i-1]\ge4\delta\).

For each run \(R\), store its boundaries and its \emph{base} \(b_R=S_0[\min R]\).
Represent the current prefix sums as
\( S[i]=b_R+c_i \)
for each $i\in R$,
where \(c_i\) is the offset from the frozen base.
Since a run contains at most \(\delta\) indices, its frozen prefix sums span \(O(\delta^2)\).
Moreover, each prefix sum changes by at most \(\delta\) between rebuilds.
Consequently, \(|c_i|=O(\delta^2)\) throughout this period.

Each offset therefore requires \(O(\log\delta)\) bits.
We pack all offsets into a constant number of machine words,
using fixed-width signed fields separated by \emph{guard bits}.
These are spare bits between adjacent fields that absorb
carries during packed arithmetic and permit independent
word-parallel comparisons; they are cleared after updates.
For sufficiently small \(\varepsilon\), all fields and
their guard bits occupy \(O(\delta\log\delta)=O(w)\) bits.
An update adds \(\pm1\) to the packed offsets at all
indices \(i\ge k\) in constant time.
A prefix-sum query reads the corresponding base and offset.

Store the run bases~$b_R$ in a dynamic fusion node \cite{patrascu14dynamic}.
For \(\operatorname{search}(q)\), find the last base
\(b_\lambda\le q\), or let \(\lambda=0\) if none exists.
Consecutive runs are separated by frozen prefix-sum
gaps of at least \(4\delta\), whereas each prefix sum
changes by at most \(\delta\) before rebuilding.
Hence, the answer lies in run \(\lambda\), run \(\lambda+1\), or at the first position of run \(\lambda+2\), 
ignoring nonexistent runs.

Examine these candidates in order.
For a candidate run \(R\), skip it if \(S[\max R]<q\),
and return \(\min R\) if \(S[\min R]\ge q\).
Otherwise, $q \in [S[\min R]..S[\max R]]$, 
so \(q-b_R=O(\delta^2)\) fits in one packed word.
Comparing all offsets of \(R\) with \(q-b_R\)
in parallel identifies the first index with \(S[i]\ge q\).
This remains correct when entries are zero: the prefix
sums are nondecreasing, and we return the first
qualifying index.

Constructing the runs and inserting their at most
\(\delta\) bases into the dynamic fusion node takes
\(O(\delta)\) time and words.
Rebuilding every \(\delta\) updates gives
\(O(1)\) amortized time per operation.
\end{proof}

\paragraph{Accelerating the two-level LPT.}
Let \(\tau\in[3..n]\), let \(w\in\Omega(\log n)\), and let the
prepared batch alphabet be \(\Gamma_B\), with
\(g=|\Gamma_B|\le\tau+1\).
For nonconstant \(\tau\), choose
\(\delta=\Theta((\log\tau)^\varepsilon)=O(w^\varepsilon)\)
for a sufficiently small constant \(\varepsilon>0\).

We retain the outer anchor tree \(\mathcal B\) and the local
bucket trees \(\mathcal B_j\) of \cref{sec:lpt}, implementing
both levels as B trees of degree \(\delta\).
An outer entry consists of the static gap preceding anchor \(j\)
and its bucket \(\mathcal B_j\); its weight is the gap length
plus the number of breakpoints in that bucket.
Each local tree orders the individual breakpoints in its bucket,
each of weight one.
Static gaps contribute no pending-letter counts.
The leaves of both levels keep \(O(\delta)\) entries in
balanced binary search trees, augmented with subtree weights and
per-letter pending counts for \(\mathtt c\in\Gamma_B\).

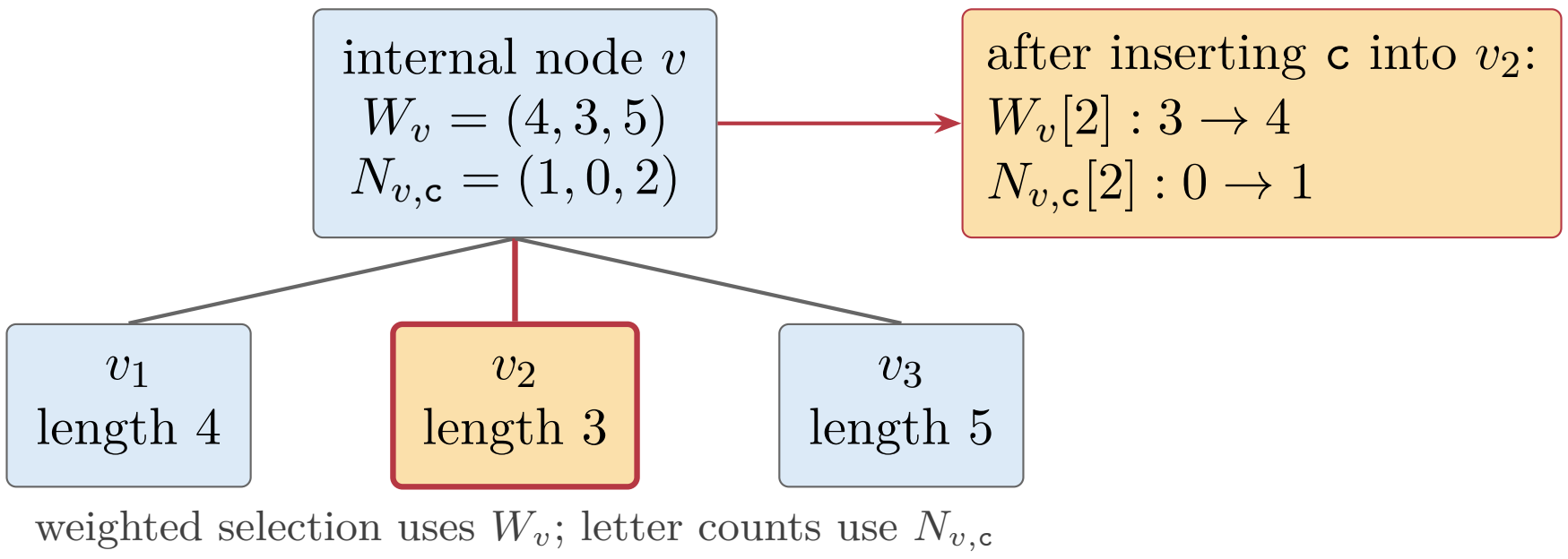
\begin{figure}[t]
\centering
\begin{tikzpicture}[x=0.95cm,y=0.85cm]
  \node[draw=black!60,rounded corners=2pt,fill=figStatic,
        align=center,inner sep=6pt] (v) at (3,3.8)
        {internal node \(v\)\\[-1pt]
         \(\displaystyle W_v=(4,3,5)\)\\[-1pt]
         \(\displaystyle N_{v,\mathtt c}=(1,0,2)\)};
  \node[draw=black!60,rounded corners=2pt,fill=figStatic,
        align=center,minimum width=18mm,minimum height=12mm] (v1)
        at (0,1.35) {\(v_1\)\\length \(4\)};
  \node[draw=figMarker,very thick,rounded corners=2pt,fill=figInsert,
        align=center,minimum width=18mm,minimum height=12mm] (v2)
        at (3,1.35) {\(v_2\)\\length \(3\)};
  \node[draw=black!60,rounded corners=2pt,fill=figStatic,
        align=center,minimum width=18mm,minimum height=12mm] (v3)
        at (6,1.35) {\(v_3\)\\length \(5\)};
  \draw[black!60,thick] (v.south) -- (v1.north);
  \draw[figMarker,very thick] (v.south) -- (v2.north);
  \draw[black!60,thick] (v.south) -- (v3.north);
  \node[draw=figMarker,fill=figInsert,rounded corners=2pt,
        align=left,inner sep=5pt] (updated) at (8.8,3.8)
        {after inserting \(\mathtt c\) into \(v_2\):\\
         \(W_v[2]:3\to4\)\\
         \(N_{v,\mathtt c}[2]:0\to1\)};
  \draw[figarrow,figMarker] (v.east) -- (updated.west);
  \node[figlabel,text=black!75] at (3,0.26)
        {weighted selection uses \(W_v\); letter counts use \(N_{v,\mathtt c}\)};
\end{tikzpicture}
\caption{Search summaries at a B-tree node.  The length weights
\(W_v[k]\) count the entire materialized sequence in child \(v_k\),
whereas \(N_{v,\mathtt c}[k]\) counts only pending occurrences of
\(\mathtt c\).  Insertion into an existing bucket changes one entry
per array along each affected path; a child split rebuilds its
parent's arrays.}
\label{fig:btree-summaries}
\end{figure}

For each internal node \(v\), with children \(v_1,\ldots,v_d\),
store the arrays
\[
\begin{aligned}
W_v[k] &= |\text{materialized sequence represented by }v_k|,\\
N_{v,\mathtt c}[k]
       &= |\text{pending occurrences of }\mathtt c\text{ in }v_k| \text{~for every~} \mathtt c\in\Gamma_B.
\end{aligned}
\]
These arrays use SSPS (\cref{lem:btree-micro}), 
padding unused entries with zeros.  
Weighted selection first searches the outer tree $\mathcal B$ and, 
when its result is a bucket, the corresponding local tree~$\mathcal B_j$.  
\Cref{fig:btree-summaries} illustrates the two arrays at an internal node of either level.

Weighted selection in the outer B tree~$\mathcal B$ locates a static gap or a bucket.
For a gap \(Z[a..b]\), 
a cut after \(t\) letters has static coordinate \(a+t\).
For a bucket \(\mathcal B_j\), 
an order-statistic query locates the boundary after the first \(t\) breakpoints.

Similarly, \(\operatorname{bcount}_{\mathtt c}(\chi)\) combines an
outer prefix count with a prefix count in the selected bucket.
Inserting \(\mathtt c\) increments the weight and the
\(\mathtt c\)-count on each affected search path.
Creating a bucket inside a static gap splits that gap in the outer
leaf but changes the total represented length by only one.

A separate searchable partial-sums tree stores \(P[\mathtt c]\), 
the number of buffered breakpoints carrying each letter \(\mathtt c\in\Gamma_B\).
Together with the static counts \(C_Z\), it supports
\[
L.\fnCount(\mathtt c)
= C_Z[\mathtt c]
  + \sum_{\substack{\mathtt d\in\Gamma_B, \mathtt d<\mathtt c}} P[\mathtt d].
                \]

We also replace the letter-keyed BST \(\mathcal D\) by a high-degree B tree equipped with SSPS,
maintaining the multiplicities \(P[\mathtt c]\) of buffered letters.

\begin{lemma}[Accelerated construction BBS]\label{lem:btree}
The dynamic part of the BBS supports
\(\myFunction{accessCut}\), \(\myFunction{access}\),
\(\operatorname{bcount}_{\mathtt{c}}\), and \(L.\fnCount(\mathtt{c})\) in
\(O(\log\tau/\log\log\tau)\) time.
Each breakpoint insertion takes the same amortized time apart from rebalancing and reconstruction of alphabet-wide summaries, 
whose \emph{total} cost during a batch of at most \(\tau\) insertions is \(O(\tau g)\).  
The structure uses \(O(\tau g)\) words.
\end{lemma}

\begin{proof}
For constant \(\tau\), both levels have constant size and can be
searched and updated directly, e.g., with plain arrays.
Otherwise, each B tree has height \(O(\log_\delta\tau)\), 
and a leaf operation takes \(O(\log\delta)\) time.  
Each query visits the outer tree $\mathcal B$ and at most one local tree $\mathcal B_j$; 
by \cref{lem:btree-micro}, their internal-node summaries take \(O(1)\) amortized time per level.
Ordinary insertions update one counter per affected node at each
level.  Creating an anchor redistributes the static gap inside
one outer leaf in \(O(\log\delta)\) time; above that leaf,
the total weight still increases by one unless a node splits.
Thus queries and ordinary insertions take
\(O(\log_\delta\tau+\log\delta) = O(\log\tau/\log\log\tau)\).

The B tree \(\mathcal D\) has at most \(g\le\tau+1\)
entries, so its operations satisfy the same time bound.

Rebalancing a leaf uses \(O(1)\) rotations per insertion, each
recomputing \(O(g)\) counters.  A B-tree split rebuilds its
parent's \(O(g)\) arrays of length \(O(\delta)\), costing
\(O(\delta g)\); the total weight and counts of the parent are
unchanged by the split, so higher ancestors need no rebuilding.
Across both tree levels, \(O(\tau)\) insertions cause
\(O(\tau/\delta)\) splits and \(O(\tau)\) leaf rotations.
Their total reconstruction cost is therefore \(O(\tau g)\).
There are \(O(\tau)\) outer entries and inner breakpoints,
with \(O(g)\) summary information per stored entry, giving
\(O(\tau g)\) words in total.
\end{proof}

Since the batch alphabet \(\Gamma_B\) is fixed,
we build a fusion tree on its sorted letters to support predecessor and equality queries.
This dictionary is used to index the static rank samples and cumulative counts of \(Z\).
A fusion tree~\cite{patrascu14dynamic} over the sorted batch alphabet answers predecessor queries in \(O(1+\log\tau/\log w)\subseteq O(\log\tau/\log\log\tau)\) time.
It can be built by successive insertion in \(O(g(1+\log\tau/\log w))\) time.
Preparing and sorting \(\Gamma_B\) takes \(O(\tau(1+\log g))\subseteq O(\tau g)\) time per batch.
Together with the sampled rank structure on \(Z\), 
this yields BBS rank operations in \(O(z/\tau+\log\tau/\log\log\tau)\) time.

The following lemma addresses the implementation of the operations on the logical edit representation of \cref{sec:inversion}; 
the factor-by-factor decoding invariant~\ref{inv:inverse-flush} is stated before \cref{lem:inverse-invariants}.

\begin{lemma}[Accelerated inversion batch]\label{lem:btree-inverse}
Let \(Z\) be the static transform of length \(z\) at the start
of an inversion batch, let \(\sigma=|\Sigma|\), and assume
\(\tau\sigma=O(n)\).
The deletion representation of \cref{sec:inversion}, including
the single virtual or temporarily substituted delimiter, supports
each destructive \(\FL\)-step in
\(
 O\!\left(\frac{z}{\tau}
          +\frac{\log\tau}{\log\log\tau}
          +\sigma\right)
\)
time and uses \(O(\tau\sigma)\) additional words.
The batch structures are initialized in \(O(z+\tau\sigma)\) time.
\end{lemma}

\begin{proof}
Rank-compress the alphabet once to \(\Sigma=[1..\sigma]\), retaining
the original letters for output, in
\(O(n(1+\log\sigma))\subseteq O(n\sigma)\) time and \(O(\sigma)\)
words.

Partition \(Z\) into \(O(\tau)\) blocks of length
\(O(\lceil z/\tau\rceil)\).  For every \(\mathtt{c}\in\Sigma\), maintain
surviving counts per block in a high-degree searchable partial-sums
tree (\cref{lem:btree-micro}).  Scanning \(Z\) initializes these
trees in \(O(z+\tau\sigma)\) time and \(O(\tau\sigma)\) words;
decrements, prefix counts, and block selection take
\(O(\log\tau/\log\log\tau)\) time.

Scan the selected block, skipping excluded coordinates
\(\mathcal E^\star\) via links between successive exclusions.
A dynamic fusion tree~\cite{patrascu14dynamic} supplies their
predecessor and rank in \(O(1+\log\tau/\log w)\) time, so one
predecessor search suffices for an \(O(z/\tau+1)\)-time scan.
A static coordinate \(h\) maps to logical position
\(h-\operatorname{rank}_{\mathcal E^\star}(h)\), adjusted by one
comparison for the delimiter, whose position is stored separately.

An \(O(\sigma)\) scan of the surviving letter totals, with the stored
delimiter-presence bit, determines \(F[i]\) and its cumulative count.
Replacing an ordinary coordinate
\(h\) by the delimiter decrements one block count and inserts
\(h\) into \(\mathcal E^\star\); an old delimiter coordinate is already
excluded and needs no second decrement.  Virtual delimiter changes
affect only constant-size state.  The claimed bounds follow.
\end{proof}

\myblock{Initialization cost}
Initializing counters and rebuilding nodes costs
\(O(\tau\sigma_\tau)\) per batch, already incurred by sampled ranks
in \cref{subsec:bbs-structure}.  Over all batches this is
\(O(n\sigma_\tau)\subseteq O(n^2/\tau)\) when
\(\tau\sigma_\tau=O(n)\), which we assume below.

The ordinary BWT uses the same BBS, so the acceleration applies
to both transforms.

\begin{theorem}\label{thm:bwt:btree}
    Let \(T\) be a text of length \(n\), and let \(\tau \in [3..n]\).
We can compute the (B)BWT in
\(
	O\!\left(
		\left(\frac{n^2}{\tau}+n\right)
		\frac{\log\tau}{\log\log\tau}
	\right)
\)
time using \(O(\tau\sigma_\tau)\) additional space, assuming \(\tau\sigma_\tau=O(n)\).
Here \(\sigma_\tau\le\min(\sigma,\tau+1)\) denotes the maximum alphabet size charged to a batch.
For inversion, we obtain the same time bound in
\(O(\tau\sigma)\) additional space under the corresponding
assumption \(\tau\sigma=O(n)\).
\end{theorem}

\begin{proof}
In a construction batch, 
\(Z\) stays fixed and at most \(\tau\) insertions are stored as breakpoints.  

We scan \(Z\) once to compute \(C_Z\) and the static rank samples: 
for each \(\mathtt c \in \Gamma_B\),
we record its prefix frequency at the \(O(\tau)\) segment boundaries.
Each scanned letter requires one predecessor query in the batch alphabet, taking \(O(1+\log\tau/\log w)\) time.

By \cref{lem:btree}, 
breakpoint operations cost \(O(\log\tau/\log\log\tau)\) amortized time, 
with rebuilding charged separately.  
Direct scans and materialization add \(O(z+\tau)\) time.
Thus, a batch of static length \(z\) costs
\[
		O\!\left(z\left(1+\frac{\log\tau}{\log w}\right)+\tau\frac{\log\tau}{\log\log\tau}+\tau\sigma_\tau\right)
	\subseteq
	O\!\left((z+\tau)\frac{\log\tau}{\log\log\tau}+\tau\sigma_\tau\right) \text{~time},
\]
since \(\log w\in\Omega(\log\log\tau)\).

Let \(z_b\) be the static transform length at the start of batch \(b\). 
Since there are \(O(n/\tau)\) batches and \(z_b\le n\), 
we have
\(
\sum_{\text{batches }b} z_b=O(n^2/\tau).
\)
The space bound follows from \cref{lem:btree}.

For inversion, \cref{lem:btree-inverse} gives
\(O(z+\tau\sigma)\) initialization time and
\(O(z/\tau+\log\tau/\log\log\tau+\sigma)\) time per step.
With at most \(\tau\) steps per batch, 
the total is \(O(n^2/\tau+n\log\tau/\log\log\tau+n\sigma)\) time.
The condition \(\tau\sigma=O(n)\) absorbs the last term,
including initial rank compression.
\end{proof}

\myblock{Refinement}
The static scans more precisely take
\(O\!\left(\frac{n^2}{\tau}(1+\frac{\log\tau}{\log w})\right)\) time.
For \(w=\Theta(\log n)\), this is
\(O\!\left(\frac{n^2}{\tau}(1+\frac{\log\tau}{\log\log n})\right)\) time,
improving on the uniform bound for small \(\tau\).
If batch-alphabet lookup is constant time, the static scans become
linear in their lengths.  
Since \(\sigma_\tau=O(\tau)\), this holds when \(\tau\in w^{O(1)}\), even for large global alphabets.
It may also hold for much larger \(\tau\), for example when \(\sigma=O(1)\).  
The next theorem uses the weaker condition \(\sigma_\tau\in w^{O(1)}\).

\begin{theorem}\label{thm:bwt:btree-small-gamma}
    Let \(\tau \in [3..n]\) and \(\tau\sigma_\tau=O(n)\).
If \(\sigma_\tau\in w^{O(1)}\), i.e.\ each batch alphabet has size \(w^{O(1)}\),
we can compute the (B)BWT in
\(
	O\!\left(
		\frac{n^2}{\tau}
		+
		\frac{n\log\tau}{\log\log\tau}
	\right)
\)
time using \(O(\tau\sigma_\tau)\) additional space.
For inversion, we obtain the same bounds by replacing \(\sigma_\tau\) with \(\sigma\), under the corresponding hypothesis \(\sigma\in w^{O(1)}\).
\end{theorem}

\begin{proof}
With \(\sigma_\tau\in w^{O(1)}\), a dynamic fusion tree answers batch-alphabet predecessor and equality queries in \(O(1)\) time~\cite{patrascu14dynamic}.
Key preparation takes \(O(\tau(1+\log\sigma_\tau))\subseteq O(\tau\sigma_\tau)\) time.  
Each scanned letter of \(Z\) can then be located in the batch alphabet in constant time, so building \(C_Z\) and rank samples takes \(O(z)\).
The at most \(\tau\) dynamic operations cost \(O(\log\tau/\log\log\tau)\) each by \cref{lem:btree}.
The batch cost is therefore
\(
O\!\left(z+\tau\frac{\log\tau}{\log\log\tau}+\tau\sigma_\tau\right) 
\)
time.
Summing over batches and using \(\tau\sigma_\tau=O(n)\) gives
\(
	O\!\left(
		\frac{n^2}{\tau}
		+
		\frac{n\log\tau}{\log\log\tau}
	\right)
\)
time.
\end{proof}

\myblock{Where the second term matters}
The additive term \(n\log\tau/\log\log\tau\) in the time bound of \cref{thm:bwt:btree-small-gamma} is dominated by \(n^2/\tau\) unless \(\tau=\omega(n\log\log\tau/\log\tau)\); 
thus the primary gain over \cref{thm:bbwt:basicTradeoff} is in the first term.

For \(\tau\le\sqrt n\), 
the at most \(\tau\) pending breakpoints can instead be maintained in a sorted array.
Each insertion or pending-breakpoint query takes \(O(\tau)\) time, contributing \(O(n\tau) \subseteq O(n^2/\tau)\) overall.
Thus, balanced search trees become necessary for our time bounds only when \(\tau > \sqrt n\).

Our construction uses \(O(\tau\sigma_\tau)\) additional words, 
which is sublinear if \(\sigma_\tau=o(n/\tau)\).
This condition restricts the number of distinct letters within each batch, 
but not necessarily the global alphabet size \(\sigma\), 
since different batches may contain different letters.

\section{Pointer-Machine Interpretation}
\label{sec:pointer-machine}
The basic trade-off does not require constant-time indexed access to the working string~$X$.  
On a pointer machine, we instead assume that the input is supplied as an ordered tree with \(n\) writable leaves,
storing the text in left-to-right order.  
The leaves form a doubly linked list, and pointers to its ends are available.  
The given tree and leaf links are part of the input/output representation, 
not additional workspace; 
the output overwrites the leaf labels without changing their order.  
We allow constant-time comparisons and arithmetic on positional counters, 
but no word-parallel operations.

All traversals of the working string can then use \emph{fingers} (pointers to leaves).  
Duval's factorization keeps fingers (represented by the state $\mathcal Q$) at its active positions; 
resetting one finger to another takes constant time, 
while advancing along the leaf list retains the standard linear-time bound.  
Each factor is reversed by swapping labels from its two ends.  
During a batch, pointers to the sampled segment boundaries locate the starts of static scans, 
and the BBS trees locate the segment containing a queried logical position.
Materialization traverses the leaves right to left for construction
and left to right for inversion.  
None of these operations needs to reach a leaf by its numerical index.

Consequently, the comparison-based construction and inversion bounds of \cref{thm:bbwt:basicTradeoff} remain valid, 
with the same \(O(\tau\sigma_\tau)\) and \(O(\tau\sigma)\) additional space bounds, respectively,
recalling that the space is relative to the \emph{given} linked-leaf representation.

\bibliographystyle{plainnat}

\begin{thebibliography}{16}
\providecommand{\natexlab}[1]{#1}
\providecommand{\url}[1]{\texttt{#1}}
\expandafter\ifx\csname urlstyle\endcsname\relax
  \providecommand{\doi}[1]{doi: #1}\else
  \providecommand{\doi}{doi: \begingroup \urlstyle{rm}\Url}\fi

\bibitem[Bannai et~al.(2024)Bannai, K{\"{a}}rkk{\"{a}}inen, K{\"{o}}ppl, and Pi\k{a}tkowski]{bannai24constructing}
Hideo Bannai, Juha K{\"{a}}rkk{\"{a}}inen, Dominik K{\"{o}}ppl, and Marcin Pi\k{a}tkowski.
\newblock Constructing and indexing the bijective and extended {Burrows}--{Wheeler} transform.
\newblock \emph{Inf. Comput.}, 297\penalty0 (105153):\penalty0 1--30, 2024.
\newblock \doi{10.1016/j.ic.2024.105153}.

\bibitem[Bannai et~al.(2025)Bannai, K{\"{o}}ppl, and Lipt{\'{a}}k]{bannai25survey}
Hideo Bannai, Dominik K{\"{o}}ppl, and Zsuzsanna Lipt{\'{a}}k.
\newblock A survey of the bijective burrows-wheeler transform.
\newblock In \emph{The Expanding World of Compressed Data: A Festschrift for {Giovanni} {Manzini}'s 60th Birthday}, volume 131 of \emph{OASIcs}, pages 2:1--2:26, 2025.
\newblock \doi{10.4230/OASIcs.Manzini.2}.

\bibitem[Burrows and Wheeler(1994)]{burrows94bwt}
Michael Burrows and David~J. Wheeler.
\newblock A block sorting lossless data compression algorithm.
\newblock Technical Report 124, Digital Equipment Corporation, Palo Alto, California, 1994.

\bibitem[Crochemore et~al.(2015)Crochemore, Grossi, K{\"{a}}rkk{\"{a}}inen, and Landau]{crochemore15bwt}
Maxime Crochemore, Roberto Grossi, Juha K{\"{a}}rkk{\"{a}}inen, and Gad~M. Landau.
\newblock Computing the {Burrows--Wheeler} transform in place and in small space.
\newblock \emph{J. Discrete Algorithms}, 32:\penalty0 44--52, 2015.
\newblock \doi{10.1016/j.jda.2015.01.004}.

\bibitem[Crowley(1998)]{crowley98textsequences}
Charles Crowley.
\newblock Data structures for text sequences.
\newblock Technical report, University of New Mexico, 1998.
\newblock URL \url{https://www.cs.unm.edu/~crowley/papers/sds.pdf}.

\bibitem[da~Louza et~al.(2017)da~Louza, Gagie, and Telles]{louza17bwt}
Felipe~Alves da~Louza, Travis Gagie, and Guilherme~P. Telles.
\newblock {Burrows--Wheeler} transform and {LCP} array construction in constant space.
\newblock \emph{J. Discrete Algorithms}, 42:\penalty0 14--22, 2017.
\newblock \doi{10.1016/j.jda.2016.11.003}.

\bibitem[I et~al.(2025)I, K\"{o}ppl, Sakamoto, and Yamaguchi]{i25spaceefficient}
Tomohiro I, Dominik K\"{o}ppl, Hiroshi Sakamoto, and Sohei Yamaguchi.
\newblock Space-efficient {B} trees via load-balancing.
\newblock \emph{Theory of Computing Systems}, 69\penalty0 (32):\penalty0 1--26, 2025.
\newblock \doi{10.1007/s00224-025-10238-7}.

\bibitem[Kempa and Kociumaka(2019)]{kempa19synchronizing}
Dominik Kempa and Tomasz Kociumaka.
\newblock String synchronizing sets: sublinear-time {BWT} construction and optimal {LCE} data structure.
\newblock In \emph{Proc.\ STOC}, pages 756--767, 2019.
\newblock \doi{10.1145/3313276.3316368}.

\bibitem[K{\"{o}}ppl et~al.(2020)K{\"{o}}ppl, Hashimoto, Hendrian, and Shinohara]{koppl20inplace}
Dominik K{\"{o}}ppl, Daiki Hashimoto, Diptarama Hendrian, and Ayumi Shinohara.
\newblock In-place bijective {Burrows}--{Wheeler} transforms.
\newblock In \emph{Proc.\ CPM}, volume 161 of \emph{LIPIcs}, pages 21:1--21:15, 2020.
\newblock \doi{10.4230/LIPIcs.CPM.2020.21}.

\bibitem[Mantaci et~al.(2007)Mantaci, Restivo, Rosone, and Sciortino]{mantaci07ebwt}
Sabrina Mantaci, Antonio Restivo, Giovanna Rosone, and Marinella Sciortino.
\newblock An extension of the {Burrows--Wheeler} transform.
\newblock \emph{Theor. Comput. Sci.}, 387\penalty0 (3):\penalty0 298--312, 2007.
\newblock \doi{10.1016/j.tcs.2007.07.014}.

\bibitem[Newman(2012)]{newman12design}
William Newman.
\newblock Design case study: the bravo text editor.
\newblock \emph{Interactions}, 19\penalty0 (1):\penalty0 75--80, 2012.
\newblock \doi{10.1145/2065327.2065342}.
\newblock URL \url{https://doi.org/10.1145/2065327.2065342}.

\bibitem[Nong et~al.(2009)Nong, Zhang, and Chan]{nong09sais}
Ge~Nong, Sen Zhang, and Wai~Hong Chan.
\newblock Linear suffix array construction by almost pure induced-sorting.
\newblock In \emph{Proc.\ DCC}, pages 193--202, 2009.
\newblock \doi{10.1109/DCC.2009.42}.

\bibitem[Olbrich(2025)]{olbrich25fast}
Jannik Olbrich.
\newblock Fast and memory-efficient {BWT} construction of repetitive texts using {Lyndon} grammars.
\newblock In \emph{Proc.\ ESA}, volume 351 of \emph{LIPIcs}, pages 60:1--60:19, 2025.
\newblock \doi{10.4230/LIPICS.ESA.2025.60}.

\bibitem[Olbrich et~al.(2024)Olbrich, Ohlebusch, and B{\"{u}}chler]{olbrich24generic}
Jannik Olbrich, Enno Ohlebusch, and Thomas B{\"{u}}chler.
\newblock Generic non-recursive suffix array construction.
\newblock \emph{{ACM} Trans. Algorithms}, 20\penalty0 (2):\penalty0 18, 2024.
\newblock \doi{10.1145/3641854}.

\bibitem[P{\u{a}}tra{\c{s}}cu and Demaine(2004)]{patrascu04tight}
Mihai P{\u{a}}tra{\c{s}}cu and Erik~D. Demaine.
\newblock Tight bounds for the partial-sums problem.
\newblock In \emph{Proc.\ SODA}, pages 20--29, 2004.
\newblock URL \url{http://dl.acm.org/citation.cfm?id=982792.982796}.

\bibitem[Patrascu and Thorup(2014)]{patrascu14dynamic}
Mihai Patrascu and Mikkel Thorup.
\newblock Dynamic integer sets with optimal rank, select, and predecessor search.
\newblock In \emph{Proc.\ FOCS}, pages 166--175, 2014.
\newblock \doi{10.1109/FOCS.2014.26}.

\end{thebibliography}

\end{document}